\documentclass[prd,nofootinbib,floats,superscriptaddress,eqsecnum,tightenlines,11pt]{revtex4}

\usepackage{hyperref}
\usepackage{graphicx}
\usepackage{amsmath,amssymb,amsfonts,amsthm,latexsym,stmaryrd,bbm}
\usepackage{marginnote}
\usepackage{color}

\def\be{\begin{equation}}
\def\ee{\end{equation}}
\def\ba{\begin{eqnarray}}
\def\ea{\end{eqnarray}}
\def\bas{\begin{subequations}\begin{eqnarray}}
\def\eas{\end{eqnarray}\end{subequations}}

\def\SU{\text{SU}}

\newcommand{\mat} [2] {\left ( \begin{array}{#1}#2\end{array} \right ) }

\begin{document}

\title{Analytic continuation of the rotating black hole state counting}

\author{Jibril Ben Achour}
\email{jibrilbenachour@gmail.com}
\affiliation{Departement of Physics, Center for Field Theory and Particles Physics, Fudan University, 20433 Shanghai, China}

\author{Karim Noui}
\email{karim.noui@lmpt.univ-tours.fr}
\affiliation{F\'ed\'eration Denis Poisson, Laboratoire de Math\'ematiques et Physique Th\'eorique (UMR 7350), Universit\'e Fran\c cois Rabelais, Parc de Grandmont, 37200 Tours, France}
\affiliation{Laboratoire APC -- Astroparticule et Cosmologie, Universit\'e Paris Diderot Paris 7, 75013 Paris, France}

\author{Alejandro Perez}
\email{perez@cpt.univ-mrs.fr}
\affiliation{Centre de Physique ThŽ\'eorique (UMR 7332) , Aix Marseille Universit\'e and Universit\'e de Toulon, 13288 Marseille, France}

\begin{abstract}
In loop quantum gravity, a spherical black hole can be described in terms of a Chern-Simons theory on a punctured 2-sphere. The sphere represents the horizon. The punctures are the edges of spin-networks in the bulk which cross the horizon and carry quanta of area. One can generalize this construction and model a rotating black hole by adding an extra puncture colored with the angular momentum $J$ in the 2-sphere. 
We compute the entropy of rotating black holes in this model and study
its  semi-classical limit. 
After performing an analytic continuation which sends the Barbero-Immirzi parameter to $\gamma = \pm i$, we show that the leading order term in the semi-classical expansion of the entropy reproduces the Bekenstein-Hawking  law independently of the value of $J$. 
\end{abstract}

\maketitle

%\tableofcontentss

\section{Introduction}
Loop Quantum Gravity (LQG) provides a microscopic explanation of the entropy 
of black holes.  Different approaches have been developed the last twenty years
within the framework of LQG  to propose a fundamental description of (mainly) spherical black holes
with no angular momentum (see \cite{Alereview} for a recent review). 
The very first model was introduced in \cite{Rovelli} where it was shown that 
intertwiners could be the good notion to describe microstates of black holes. The introduction
 of isolated horizons \cite{Ash0, Ash1, Ash2, Ash3} allowed for a description of  the black hole
microstates in terms of a $U(1)$ Chern-Simons theory \cite{Ash4,Ash5} on a 2-sphere representing the black hole
horizon.  Later, it was shown in \cite{BHentropy8, BHentropy6} that black holes can be treated in terms of an $SU(2)$ Chern-Simons theory which 
further clarified the relation between the microstates of the black holes and the spin-network states in the bulk outside
the horizon. Furthermore, in that picture, counting the number of microstates is remarkably simple and relies (at the semi-classical limit) on counting $SU(2)$ intertwiners between unitary irreducible representations \cite{BHentropy7,kaulma1,kaulma2,kaulma3}. 
While  the understanding of the black hole microstates has constantly improved in LQG,  many different techniques to compute their number  were developed \cite{counting1,counting2,counting3,counting4,counting5,BHentropy3,Liv1,Bianchi}.
However, all these models have in
common that counting the microstates leads to recovering the Bekenstein-Hawking law \cite{Bekenstein,Hawking} provided that the Barbero-Immirzi parameter $\gamma$ is fixed to a special value. 
The fact that $\gamma$, which is 
classically irrelevant, plays such an important role has raised some doubts and criticisms. In particular,  one natural 
question would be to ask whether or not the value of $\gamma$ changes for rotating black holes in these 
different approaches to recover the Bekenstein-Hawking law. 

For reasons that will be explained below the rotating case has shown to be rather elusive. To our knowledge, 
the first attempt to model rotating black holes in LQG was proposed in \cite{Kras1}. 
Nonetheless, at the classical level, the concept of spherically isolated horizon was quickly generalized to include rotation \cite{Ash9,Ash10}, and also distortion \cite{Ash7} or couplings to fields \cite{Ash8}. Further investigations were carried out in order to relate the rotating non expanding horizon concept to the well known Kerr geometry in \cite{Lew1}. Although a quantization of the rotating isolated horizon following the technics used for the static spherically isolated horizon was explored in \cite{Ash10}, the problem is still open. In some of these models \cite{Ash9,Ash10}, it was argued that the resulting entropy matches the Bekenstein-Hawking area law for the same peculiar value of the Immirizi parameter $\gamma$ appearing in the non rotating case. However, the isolated horizon definition breaks  the diffeomorphism symmetry on the horizon in the rotating case, which ,in the quantum model, makes the number of states uncountably large.
Another attempt to describe microstates and to compute the entropy of a rotating isolated horizon was presented in \cite{Boj1}, where a new angular momentum operator was introduced. A different effective approach to compute the entropy of a rotating black hole based on an analogy with the physics of polymer chain was also proposed in \cite{Bianchi}.  
There is a way in which one can resolve the conflict between the isolated horizon boundary condition and diffeomorphisms in the rotating case \cite{Ale1}. The idea consists of using an enlargement of the field content on the horizon, which account for the angular momentum, in order to restore the diffeomorphism invariance. The resulting model and its quantization was introduced in \cite{Ale2}. The important feature of this model is that the fundamental geometric  degrees of freedom at the horizon are described in terms of an $SU(2)$ Chern-Simons theory which is locally indistinguishable from the theory describing the non-rotating BHs. For that reason this model is the only quantum mechanical description that is compatible, on the one hand, with diffeomorphim invariance and, on the other hand, with the gauge properties of bulk quantum states of geometry represented by $SU(2)$ spin network states.

However, a problem remains. Indeed the state counting for the model \cite{Ale2} (revisited in Section \ref{gamma real} in this paper) but also for the models  \cite{Liv1,Bianchi}  (which to our knowledge are the only models compatible with the $SU(2)$ gauge symmetry in the bulk) yields an entropy with an {\bf explicit} dependence on the angular momentum $J$ of the black hole.  This result is incompatible with the Bekenstein-Hawking area law. In this paper we will show that this tension seem to disappear when we implement the counting after analytically continuing the model \cite{Ale2} to $\gamma=\pm i$. 
Such an analytic continuation (suggesting a link between standard LQG and a self dual version of the quantum theory) has been put forward recently in the context of black holes thermodynamics \cite{Noui0,Noui1,Noui2,radiation, Pranz1, Pranz2, Pranz3, BTZ, Yasha1, Yasha2, Muxin,JBA1}. This idea has been investigated further in the context of spinfoams \cite{BN}, and in $2+1$ dimensional LQG \cite{G2, G3, J1, J2}. 
It was first shown that 
the black hole entropy, which is a function of $\gamma$, can be analytically continued to $\gamma=\pm i$ 
in a consistent way \cite{Noui1, Noui2}. Then, remarkably, the black hole entropy reproduces exactly the semi-classical area law
with the right famous $1/4$ factor.  The aim of this paper is to show that this construction works for rotating 
black holes as well.

For this purpose, we consider, as we said above, the model \cite{Ale2} of a rotating black hole in LQG. In this approach, the black hole is still
described in terms of an $SU(2)$ Chern-Simons theory on a punctured 2-sphere. The 2-sphere represents the black 
hole horizon at a given time. All the punctures but one carry as usual quanta of area which add themselves 
to form the horizon area
$a_H$. These punctures are colored with a spin $j_l$. 
The remaining puncture carries the angular momentum of the black hole which is therefore colored with the black hole
spin $J$. One can easily adapt the techniques developed for spherical black holes to compute the number of microstates
when $J$ is non-vanishing. To simplify the calculation, we focus on the so-called one-color black holes where
$j_l=j$ for any puncture which carries  quanta of area. The main result of this article would not be affected if we have
studied the most general situation.  Then, we compute the black hole entropy $S_{BH}$ 
when $\gamma$ is real and show that to
recover the Bekenstein-Hawking law one has to fix $\gamma$ to a value which is $J$-dependent. Such a condition
is of course physically non-acceptable. However, when we perform the analytic continuation to $\gamma=\pm i$ 
following the techniques developed in \cite{Noui2}, we show that the leading order term of the black hole entropy reproduces exactly the semi-classical
law whatever the value of $J$ is. More precisely, we show that \eqref{a and J}
\begin{eqnarray}
\exp S_{BH} \simeq P(a_H,J) \exp \left( \frac{a_H}{4\ell_p^2} + i2\pi J \right)
\end{eqnarray}
where $\ell_p$ is the Planck length and $P$ is a function that is responsible for quantum corrections to the entropy.
As $J$ is an integer, it does not affect the entropy of the black hole at the semi-classical
limit. There is a clear disentanglement between the angular momentum $J$ and the black hole area $a_H$.  

The paper is organized as follows. We start in Section \ref{gamma complex} with the calculation of the rotating black
hole entropy when $\gamma=\pm i$. We define the rules for the analytic continuation, then we compute the number
of microstates and we study its semi-classical limit using the saddle point approximation. 
To compare with what happens when $\gamma$ is real,  we compute the entropy of the black hole
in that case in Section \ref{gamma real}. We show that, contrary to the complex black hole case, the semi-classical
analysis does not allow us to recover, in a physically consistent way, the expected result. We close the article with a
short discussion in Section IV.

\section{Entropy of a rotating black hole}
\label{gamma complex}

In this section, we extend the proposal of \cite{Noui2} to compute the  entropy of a rotating black hole and
we study its semi-classical behavior. In a first part, we define the rules of the analytic continuation. Then, in
a second part,  we construct the entropy of the black hole with $\gamma=\pm i$.  
Finally, we show that the leading order term in the semi-classical
expansion of the entropy reproduces the Bekenstein-Hawking  law for any value 
of the black hole angular momentum $J$.

\subsection{Rules for the analytic continuation}

One crucial condition for the analytic continuation to $\gamma=\pm i$ to be consistent is that the area eigenvalues 
remain real and positive. In other words, the area operator remains  self-adjoint after the analytic continuation process. 
This condition has been analyzed in
\cite{Noui2} and  leads to the fact that the punctures associated to the horizon are no more colored with discrete spins
but with continuous real numbers. Thus, everything happens as if the discrete finite dimensional $SU(2)$ irreducible representations  are turned into continuous  infinite dimensional representations of $SU(1,1)$ (see \cite{J1,J2} for a 
very similar phenomenon in 2+1 dimensional gravity). This can be straightforwardly seen  from the expression of the area of the black hole horizon $a_H$ as a sum of its fundamental excitations $a_l$:
\begin{align}\label{area}
a_H = \sum_{\ell} a_l \qquad \text{with} \qquad
a_l= 8\pi \ell_p^2 \gamma \sqrt{j_l(j_l+1)} \; .
\end{align} 
Indeed, imposing $\gamma=\pm i$ keeps the area real and positive only
if the dimension $d_l=2j_l+1$ associated to each spin $j_l$ satisfies the condition $d_l^2 <1$. Solving this condition  depends on whether  $d_l^2 <0$ or $0<d_l^2<1$. The latter case is not relevant for our purposes and thus we focus on 
the former case. If $d_l^2 \leq 0$, then there exists  $s_l \in \mathbb R$ such that $d_l=is_l$. Hence, the spin becomes a complex number  and the area remains real according to
\begin{eqnarray}\label{discrete to continuous}
 j_l \mapsto \frac{1}{2}(-1+is_l) \, , \qquad a_l \mapsto 4 \pi \ell_p^2 \sqrt{s_l^2 + 1} \, ,
\end{eqnarray}
but the area has now a continuous spectrum. From the point of view of group representations theory, 
the mapping \eqref{discrete to continuous} is
interpreted as the mapping from unitary irreducible representations of $SU(2)$ into those of $SU(1,1)$ 
which belong to the continuous series. Such an interpretation is consistent with the observation that, after implementing the anayltic continuation, ie $\gamma = \pm i$ and $j = \frac{1}{2} (- 1 + i s) $, the level $k$ of the Chern-Simons
theory describing the non rotating black hole has to becomes purely imaginary \cite{Noui2}. Indeed having a purely imaginary (or complex) level  
is intimately related to the quantization of Chern-Simons theory with a non-compact gauge group. On the contrary, 
it is very well-known that quantizing the theory with a compact gauge group leads to a discrete level 
\cite{Witten1, Witten2}. 

\medskip

Now we want  to extend these rules to the case of rotating black holes whose angular momentum is denoted 
$J$. Rotating black
holes were modeled in terms of an $SU(2)$ Chern-Simons theory on a punctured 2-sphere with an extra puncture 
colored with the spin $J$ \cite{Ale2}. This is the main difference with the description of the spherical black hole. Another difference is that the level $k$ depends now not only on $a_H$ and $\gamma$ but also on $J$  and a novel free 
parameter $\bar{\gamma}$. However,  the definition of $k$ is ambiguous as it is illustrated 
by the presence of an extra parameter $\bar{\gamma}$\footnote{In \cite{Ale2},  one makes use of this ambiguity 
and fixes the free parameter
$\bar{\gamma} $ in such a way that the level $k$ becomes  independent of the Barbero-Immirzi parameter 
$\gamma$ and it vanishes for extremal black holes. However, one could also find other arguments to fix these ambiguities.}.
For this reason, we will not pay too much attention on the exact expression of $k$ when we perform the analytic
continuation. We will only assume that $k$ becomes complex when $\gamma=\pm i$: when $J=0$, $k$ is a pure imaginary
number \cite{Noui2}, and we expect it to be complex when $J \neq 0$. Concerning the angular momentum $J$ itself, 
it is supposed to remain integer when $\gamma = \pm i$, which is natural because an angular
momentum $J$ is intimately related to an $SO(3)$ symmetry. 
As a consequence, the analytic continuation of the rotating black hole entropy is performed according   
to the following rules:
\begin{align}\label{continuation rules}
\gamma = \pm i  \, , \quad  d_l = i s_l \quad  \text{with} \quad s_l \in \mathbb{R} \, , \quad k \in \mathbb{C}  \,, \quad  
J \in \mathbb N \, .
\end{align} 
Let us emphasize again that the explicit form of $k$ is not needed for our purposes, but it becomes large at the
semi-classical limit. 

\subsection{Number of microstates}

We start with the Verlinde formula which gives the dimension of the  Chern-Simons theory Hilbert space on a punctured
2-sphere whose gauge group is $SU(2)$ and the level is $k$. When the sphere is punctured by $n$ links colored with $j_l$ and one extra link colored with $J$, this 
number is given by
\begin{align}\label{Verlinde}
N_k(J, j_l; n) \equiv  \frac{2}{k+2} \sum^{k+1}_{d=1} \sin{(\frac{\pi d}{k+2})} \sin{(\frac{\pi d d_J}{k+2})} \prod^{n}_{l=1} \frac{\sin{(\frac{\pi }{k+2} d } d_{l})}{\sin{(\frac{\pi }{k+2} d)}} 
\end{align}
where $d_J = 2J +1$ and $d_l = 2j_l +1$ are the usual dimensions  of $SU(2)$ unitary representations. This 
well-known formula allows us to compute the number of (real) black holes microstates when the condition 
\eqref{area} is imposed \cite{BHentropy7} which will be done in Section III.

\medskip

When $\gamma$ is complex, $k$ becomes also complex and then the Verlinde formula above does not make sense
anymore. However, following  the same strategy as in the non-rotating case \cite{Noui2}, it is possible to reformulate \eqref{Verlinde} as an integral on the complex plane in such a way that the analytic continuation \eqref{continuation rules}
is well-defined. After an analysis very similar to \cite{Noui2}, we show that \eqref{Verlinde} transforms to 
\begin{align}
\label{dimf}
N_{\infty}(J, s_l; n)  & = \frac{1}{i\pi} \oint_{\mathcal{C}} dz \; \sinh{z} \; \sinh{(d_{J} z)} \; \prod^{n}_{l=1} \frac{\sinh{(i s_l z)}}{\sinh{z}} 
\end{align}
when (the real and imaginary parts of) $k$ becomes large, as it is required at the semi-classical limit.
The contour $\cal C$ encloses the point $z=i\pi$ on the imaginary axe as shown in \cite{Noui2}.
We consider this function as the definition of 
the Verlinde formula when $\gamma=\pm i$  in the large $k$ limit. 

Let us remark that, contrary to the original formula \eqref{Verlinde}, there is no reason to expect that \eqref{dimf} produces a non-negative 
natural number which can be naturally interpreted as a dimension of some Hilbert space. In general, \eqref{dimf}
defines a complex number, and then its interpretation in terms of a number or even a density of  black holes microstates
is not obvious.  However, we found in \cite{Noui2} a selection rule such that \eqref{dimf} is real. 
We expect such conditions to exist
also when there is a non-vanishing angular momentum and we are going to show 
in the section \ref{section asymptotic} below 
that this is indeed the case
at least at the semi-classical limit. 
We will also show 
that the semi-classical expansion of \eqref{dimf}  reproduces at the leading order 
the Bekenstein-Hawking  law.

\subsection{Semi-classical limit and Bekenstein-Hawking law}
\label{section asymptotic}

This section is devoted to study the asymptotic behavior of \eqref{dimf} at the semi-classical limit. 
To simplify this analysis, we restrict
our study to the one-color model where the punctures are all colored with the same representation $s_l=s$. 
In this model, the black hole horizon area $a_H$ and the angular momentum
$J$ are given by
\begin{eqnarray}\label{aH and J}
a_H \, = \, 4\pi \ell_p^2 n s \,  \qquad \text{and} \qquad 
J \, = \, \frac{1}{2} a_\star  ns
\,,
\end{eqnarray}
where $a_\star\equiv 8\pi J \ell_p^2/a_H \in [0,1]$ ($a_\star=1$ for an extremal black hole). 
With this simplification, \eqref{dimf} reduces to the following integral
\begin{eqnarray}
\label{dimff}
{\cal I}_n(J, s)   \equiv  \frac{1}{i\pi} \oint_{\mathcal{C}} dz \; \sinh{z} \; \sinh{(d_{J} z)} \; \left( \frac{\sinh{(i s z)}}{\sinh{z}} \right)^n \, .
\end{eqnarray}
As it was shown in \cite{Noui2} when $J=0$, the semi-classical limit corresponds to $n$ and $s$ large (this relies on a statistical analysis in the canonical ensemble using the quasi-local approach \cite{GhoshAle1,GhoshAle2,GhoshAle3}). Introducing an
angular momentum does not change this condition, hence we are going to compute the asymptotic expansion of 
\eqref{dimff} when $n \gg 1$ and $s \gg1$. 

To apply the saddle point approximation, it is useful to write \eqref{dimff} as follows
\begin{eqnarray}\label{calI}
{\cal I}_n(J, s)   =  \oint_{\mathcal{C}} dz \; \mu_+(z) \, \exp \left[ n {\Phi}_+ (z)\right] + 
\oint_{\mathcal{C}} dz \; \mu_-(z) \, \exp \left[ n {\Phi}_- (z)\right]
\end{eqnarray}
with the measures $\mu_\pm$ and the ``actions" ${\Phi}_\pm$ defined by
\begin{eqnarray}
\mu_\pm(z) \equiv \pm \frac{1}{2i\pi} e^{\pm z}\sinh z \, \qquad \text{and} \qquad
{\Phi}_\pm(z) \equiv \pm a_\star  s z + \ln \left( \frac{\sinh(isz)}{\sinh z} \right) \, .
\end{eqnarray}
The asymptotic of each integral in the sum \eqref{calI} is evaluated by the saddle point approximation.
This method relies on the computation of the critical points $z_\pm$ of the functions ${\Phi}_\pm$ which are solutions of
the equations
\begin{eqnarray}\label{critical}
{\Phi}'_\pm(z_\pm)=0 \qquad \text{with} \qquad
{\Phi}'_\pm(z) =  \frac{s}{\tan(sz)} - \frac{1}{\tanh z} \pm a_\star   s \, .
\end{eqnarray} 
The location of the critical points was precisely analyzed  when $a_\star=0$ in \cite{Noui2}. The generalization to 
$a_\star \neq 0$ is immediate. Obviously, there are no exact solutions to the equation \eqref{critical}. 
However, one easily sees that there is an infinite number 
of solutions located on the real axe and an infinite number of solutions located at the vicinity of the points $z_p\equiv p i\pi$ 
($p\in \mathbb Z$) when $s\gg 1$. As in the case $J=0$ \cite{Noui2}, only the critical point located at the vicinity of $i\pi$ 
contributes to the semi-classical limit of \eqref{calI}. 
If we denote this critical point by $z_\pm = i\pi + \epsilon_\pm$, we easily find that
\begin{eqnarray}
\epsilon_\pm \, =  \frac{1}{s} \left( \frac{1}{ \pm  a_\star  -i} + o(1)\right) 
\, = \, \pm \frac{1}{s} \left( \frac{e^{\pm i \varphi}}{\sqrt{1+ a_\star^2 }} + o(1)\right) \,  \quad \text{with} \quad
\tan \varphi \equiv \frac{1}{a_\star} \, .\label{phi}
\end{eqnarray}
We used the notation $o(x)$  for any function $o(x) \ll x$.
Hence, the evaluation of the actions $\Phi_\pm$ at the critical points $z_\pm$ gives
\begin{eqnarray}
&&{\Phi}_\pm(z_\pm)  =  {\Phi}_\pm^0 + {\Phi}_\pm^1 \quad 
\text{with} \quad {\Phi}_\pm^0 \equiv s \pi (1 \pm i a_\star) \quad \text{and}  \label{leading}\\
 && {\Phi}_\pm^1 \equiv  \log s  +  1 
\pm i (\frac{\pi}{2} - \varphi) - i \frac{\pi}{2} 
+\log\sqrt{1+ a_\star^2} + 
o(1)  \,.\label{subleading}
\end{eqnarray}
Only the leading order term in this expansion \eqref{leading} 
will contribute to the leading order
term of the semi-classical expansion of the entropy. We  focus mainly on this leading order term and we will shortly discuss the sub-leading corrections \eqref{subleading} latter. As a consequence, the leading order term in the semi-classical expansion of the entropy $S_{BH}$ is
 given by the leading order term of the function
\begin{eqnarray}
 {\cal I}_n^0(J, s)   \equiv  
 \mu_+(z_+) \exp(n{\Phi}_+^0) + \mu_-(z_-) \exp(n{\Phi}_-^0)  \, .
\end{eqnarray}
Using \eqref{aH and J}, we show immediately  that
\begin{eqnarray}\label{a and J}
\exp(n{\Phi}_\pm^0) \; = \; \exp(\pi n s \pm i \pi ns a_\star) \; = \; \exp\left(\frac{a_H}{4\ell_p^2} \pm i 2\pi J \right)
\end{eqnarray}
where the phase disappears because $J$ is an integer.
Hence, the leading order term in the semi-classical expansion of the entropy is 
\begin{eqnarray}
S_{BH} \, = \, \frac{a_H}{4\ell_p^2}  + o\left( \frac{a_H}{\ell_p^2}\right) \, .
\end{eqnarray}
We recover the Bekenstein-Hawking law whatever the value of  the angular momentum $J$ is. 

\medskip

We can make the analysis of the semi-classical behavior  more precise
pushing further the saddle point approximation. Indeed, after a direct calculation we show that \eqref{dimf}
is equivalent, at the semi-classical limit, to
\begin{eqnarray}\label{full expansion}
 {\cal I}_n(J,s) \; \simeq \;  \sqrt{\frac{2}{n{\pi}}} \frac{1}{s^2 (1+a_\star^2)} \exp \left(\frac{a_H}{4\ell_p^2} + n [\log \sqrt{1+a_\star^2} +1 +  \log s ] \right) 
 \frac{e^{-in\varphi} + (-1)^n e^{in\varphi}}{2i}\, .
\end{eqnarray}
Hence, the analytic continuation of the number of microstates for a one-color black hole with a fixed color $s$  becomes a priori complex when $\gamma=\pm i$. 
However, one can easily cure this problem and make \eqref{full expansion} real assuming that only odd number of punctures
are allowed. This condition can be interpreted as a selection rule.  In that case, \eqref{full expansion} is real
\begin{eqnarray}\label{asymptotic 2}
{\cal I}_n(J,s) \; \simeq \; - \sqrt{\frac{2}{n{\pi}}} \frac{ \sin (n\varphi)}{s^2 (1+a_\star^2)}  \exp \left(\frac{a_H}{4\ell_p^2} + [ {\log \sqrt{1+a_\star^2}} +1 +  \log s] n\right)  
\end{eqnarray}
but it oscillates. We expect that such oscillations disappear when we compute the partition function of the black hole in the canonical ensemble. The difficulty to show precisely this is indeed the case is that
one cannot give a close formula of the partition function starting from the exact
expression \eqref{calI}. Nonetheless, to illustrate how the oscillations would disappear in a canonical ensemble,
we highly simplify the expression of \eqref{calI} and we replace it by its asymptotic expression \eqref{asymptotic 2}
where we keep only the holographic term and the oscillating term. Hence, following \cite{Noui2} we compute the partition function and we obtain
\begin{eqnarray}
\label{canonical}
{\cal Z}_{BH}(\beta) \; = \;  \sum_{n=1}^\infty  \sin(n \varphi)  \int_0^\infty \!\!ds  \exp\left( \pi  \frac{\beta_u - \beta}{\beta_u}  ns \right)\end{eqnarray}
where $\beta$ is the inverse temperature and $\beta_u$ the Unruh temperature \cite{Noui2,GhoshAle1,GhoshAle2,GhoshAle3}. In this simple example, one can easily integrate over $s$ and then perform the sum over $n$ to obtain
\begin{eqnarray}
{\cal Z}_{BH}(\beta) \; = \; \frac{\beta_u}{\pi (\beta_u - \beta)} \sum_{n=1}^\infty \frac{\sin(n \varphi) }{n} \, = \, 
\frac{\beta_u (\pi - \varphi)}{2\pi (\beta_u - \beta)}  \, .
\end{eqnarray}
As a conclusion the oscillations disappears and the partition function behaves as we expected with a divergence at
$\beta=\beta_u$. It follows from this that the canonical ensemble entropy for a macroscopic BH is given by the Bekenstein-Hawking formula at leading order. The angular momentum appears in the subleading terms \cite{Noui2,GhoshAle1,GhoshAle2,GhoshAle3}. 

Before closing this Section, let us give few remarks. First, we recall that we studied exclusively one-color black holes to make the calculations simpler and clearer. We expect the same result to hold with many-colors black hole 
exactly as in \cite{Noui2}: the leading order term of the entropy would lead to a Bekenstein-Hawking area law. The 
calculation of the canonical partition function \eqref{canonical} has been oversimplified  as well to illustrate the fact that
the oscillations we encounter in the microcanonical analysis disappear in the canonical ensemble (which is the right
ensemble to consider to show that the semi-classical limit is governed by $n$ and $s$ large). The exact partition
function can be formally written but its  calculation cannot be done exactly. 

\section{The real Barbero-Immirzi parameter case}
\label{gamma real}
To compare with the real case, we finish with the computation of the entropy  for $\gamma$ real. We are going to see that this case does not allow us to recover in a consistent way the Bekenstein-Hawking  law, contrary to what happens when $\gamma=\pm i$. A similar calculation was proposed in \cite{Liv1,Bianchi}. To make the comparison between the two situations ($\gamma$ real and $\gamma$ imaginary) simpler, we focus only on one-color black holes where all  spins are equal to the same value $j_l=j$. We denote by $d=2j+1$ the dimension of the common representation. 

When the spins are discrete, it is well-known that the Verlinde formula \eqref{Verlinde} reduces at the semi-classical
limit (with the level $k$ large) to the integral
\begin{align}
\label{dimr}
{\cal J}_n(J, d) \equiv  \int^{\pi}_0 \frac{d\theta}{2\pi} \; \sin{\theta} \sin{(d_J \theta)}  \left(  \frac{\sin{(d \theta)}}{\sin{\theta}} \right)^n \, .
\end{align}
It can be easily shown  that the Verlinde formula is nothing but the Riemann sum of \eqref{dimr}. 

The most important difference with the complex case is that the semi-classical limit is reached when the spins are small. Hence, there is no reasons here to assume that $d$ is large. On the contrary, we fix $d$ to a finite value, and
to simplify the analysis we take $d=2$ (the main results of this Section is not  affected if we consider a different value of $d$). Then, we study the semi-classical limit of
\begin{eqnarray}
{\cal J}_n(J) \equiv {\cal J}_n(J, 2) =  2^n \int^{\pi}_0  \frac{d\theta}{2\pi} \; \sin{\theta} \sin{(d_J \theta)}  \cos^n\theta \, ,
\end{eqnarray}
when $n$ and $J$ become large. To do so, we first simplify the integral which, after direct calculations,  can be reformulated 
in terms of binomial coefficients as follows
\begin{eqnarray}
\label{closedf}
{\cal J}_n( J)  = \frac{1}{2} \left[ \mat{cc}{  {n}/{2} +J \\ n}  - \mat{cc}{ {n}/{2} + J + 1 \\ n} \right]=
\frac{2(J + 1)}{ n+ 2(J+1)}  \mat{cc}{ n/2 +J\\ n}  .
\end{eqnarray}
From this expression, we see that necessarily $2J \leq n$ otherwise the number of microstates vanishes.
Hence, evaluating the semi-classical limit relies on computing the asymptotic of the binomial coefficient in \eqref{closedf}
when $n$ and $J$ become large. It is easy to show that, for any $\alpha \in [0,1]$, 
\begin{eqnarray}
\log \mat{cc}{ \alpha n \\ n} \; = \; -[\alpha \log \alpha + (1-\alpha) \log(1-\alpha)] n \, + \, o(n) \, .
\end{eqnarray}
This is an immediate consequence of the Strirling formula.
Hence, the leading-order term of the entropy for a rotating black hole is
\begin{eqnarray}
S_{BH}  =  -\frac{\alpha \log \alpha + (1-\alpha) \log(1-\alpha)}{\pi \gamma \sqrt{3}} \frac{a_H}{4\ell_p^2} + o\left( \frac{a_H}{\ell_p^2}\right) 
\quad \text{with} \quad \alpha \equiv \frac{1}{2} + \frac{J}{n} \leq 1 \, .
\end{eqnarray}
As a consequence, the requirement that $S_{BH}$ reproduces the Bekenstein-Hawking law fixes $\gamma$
to a value which depends on $J$. Of course, such a condition is not physically acceptable. 

\section{Conclusion}
In this note, we considered the model of a rotating black hole which was introduced in \cite{Ale2} in the context of Loop Quantum Gravity. Adapting the methods developed in \cite{Noui2} for spherical black holes,  we computed the analytic continuation of the number of states when the Barbero-Immirzi parameter $\gamma$ takes the complex  value $\gamma=\pm i$. We studied its semi-classical behavior and found that the leading order term of the BH entropy $S_{BH}$ reproduces exactly the expected Bekenstein-Hawking law for any values of the angular momentum $J$. This situation strongly contrasts with the case $\gamma$ real where a non trivial dependence on $J$ affects the leading order of the entropy. Let us recall
that we studied exclusively one-color black holes where the punctures which cross the horizon are colored with the same
representation. We expect, exactly as in \cite{Noui2}, that considering multi-colors black holes should not affect the main
result of the paper: their entropy should reproduce the expected law at the semi-classical limit. 

This  result   strengthens and confirms the indication that the quantum theory with (anti-)self dual variables (foreshadowed through the analytic continuation to $\gamma=\pm i$) could be better behaved that the theory defined in terms of real connection variables in view of its consistency with semiclassical expectations.  
Until now, the analytic 
continuation technique was efficient to analyze symmetry reduced models only (black holes, cosmology and 2+1
dimensional gravity). Adapting this method to have a better understanding of the full theory with (anti-)self-dual variables
is a very important problem that we hope to address in  the future.

\end{document}